\documentclass[pra,letterpaper,twocolumn,showpacs,superscriptaddress,floatfix,longbibliography]{revtex4-1}
\usepackage{graphicx,psfrag,amsmath,amssymb,amsfonts,bbm,latexsym,color,dcolumn,bm,mathbbol,mathrsfs}
\allowdisplaybreaks
\newcommand{\E}[1]{\mbox{$\mathsf E$}\left( #1 \right)}

\newcommand{\bra}[1]{\left\langle #1 \right|}
\newcommand{\ket}[1]{\left| #1 \right\rangle}
\newcommand{\abs}[1]{\left| #1 \right|}
\newcommand{\I}{\mathrm{i}}

\newcommand{\D}{\mathrm{d}}

\definecolor{myred}{RGB}{168,5,14}
\definecolor{myblue}{RGB}{13,13,255}

\begin{document}

\title{Fermi's golden rule for $N$-body systems in a blackbody
  radiation}

\author{Massimo Ostilli} \affiliation{ Departamento de F\'isica
  Te\'orica e Experimental, Universidade Federal do Rio Grande do
  Norte, Natal-RN, Brazil} \author{Carlo Presilla}
\affiliation{Dipartimento di Fisica, Sapienza Universit\`a di Roma,
  Piazzale A. Moro 2, Roma 00185, Italy} \affiliation{Istituto
  Nazionale di Fisica Nucleare, Sezione di Roma 1, Roma 00185, Italy}

\date{\today}

\begin{abstract}
  We review the calculation of Fermi's golden rule for a system of
  $N$-body dipoles, magnetic or electric, weakly interacting with a
  blackbody radiation By using the magnetic or electric field-field
  correlation function evaluated in the 1960s for blackbody radiation,
  we deduce a general formula for the transition rates and study its
  limiting, fully coherent or fully incoherent, regimes.
\end{abstract}


\maketitle

The incoherent electromagnetic (EM) radiation within a cavity at
thermal equilibrium, namely, the blackbody radiation, has, actually, a
certain degree of coherence. This is well evidenced by the analysis of
the second-order correlation function between electric or magnetic
fields reported more than 50 years ago using techniques analogous to
those employed in the theory of isotropic turbulence of an
incompressible fluid~\cite{Bourret1960,KanoWolf1962,MethaWolf1964I}.
See also \cite{Donges} for an experimental result.  Quite
surprisingly, this result has received little attention in the
literature even in dealing with problems of vast interest (see
\cite{PachonBrumer} for an exception).  Complex quantum systems,
schematized as $N$-body systems, are usually driven to thermal
equilibrium by letting them interact with blackbody radiation. In this
equilibration, often described in terms of a quantum optical master
equation~\cite{BreuerPetruccione}, the transition rates induced by the
radiation between two states of the system, as well as the spontaneous
emission contribution, describe the core processes. We do not have a
formula for these transition rates which covers the whole spectrum of
situations, from those in which the coherence properties of the
blackbody radiation are important to those in which they are
irrelevant.  The study of $N$-body systems, $N$ electric or magnetic
dipoles in the simplest case, exchanging photons with blackbody
radiation appears to be mandatory for understanding many modern
mesoscopic experiments.

In this paper, we review from the very beginning the calculation of
Fermi's golden rule for a system of $N$ dipoles, magnetic or electric,
weakly interacting with blackbody radiation. Using the magnetic or
electric field-field correlation function evaluated
in~\cite{Bourret1960,KanoWolf1962,MethaWolf1964I}, we deduce a general
formula for the transition rates and study its limiting, fully
coherent or fully incoherent, regimes.

Consider an isolated $N$-body system described by the Hermitian
Hamiltonian operator $\hat{H}$ acting on a Hilbert space $\mathscr{H}$
of dimension $M$.  For example, we have $M=2^N$ in the case of $N$
qubits.  We assume that the eigenproblem, $\hat{H} \ket{E_m} = E_m
\ket{E_m}$, has discrete, possibly degenerate, eigenvalues and that
the eigenstates $\{\ket{E_m}\}$ form an orthonormal system in
$\mathscr{H}$.  The eigenvalues are thought to be arranged in
ascending order $E_1\leq E_2 \leq \dots \leq E_M$.

We let the system interact with the EM field of blackbody radiation at
thermal equilibrium at temperature $T$. As usual, we suppose that this
interaction is sufficiently weak so that it can be tackled by a first
order perturbative analysis. We specialize the discussion to a system
consisting of $N$ charge-less spins $\bm{\sigma}_i$, $i=1,\dots,N$,
interacting with the radiation as pure magnetic dipoles.  Similar
considerations apply to spin-less charged particles interacting as
electric dipoles. The analysis is easily extended to mixed electric
and magnetic couplings. We adopt the Gaussian system of units.

Let $\mu$ be the magnetic dipole moment associated with each spin and
$\bm{r}_i$ the position vector of the $i$-th spin. The locations of
the $N$ spins are considered fixed.  Due to the interaction with the
radiation inside the cavity, the system can change its quantum state
by absorbing or emitting photons. In the semiclassical theory of
radiation, these processes are associated with the coupling of the
dipoles with, respectively, the real or the imaginary part of the
plane wave magnetic fields $\bm{B}(\bm{k})\cos(\bm{k}\cdot\bm{r}-kc
t)$.  We thus need to introduce two separate interaction operators for
each exchanged photon of wave vector $\bm{k}$,
\begin{align}
  \label{V}
  \hat{V}^\pm(\bm{k},t)= -\sum_{i=1}^{N}\mu\bm{\sigma}_i \cdot
  \frac{1}{2}\bm{B}(\bm{k}) e^{\pm\I(\bm{k}\cdot\bm{r}_i-kct)},
\end{align}
the operator with the plus sign corresponding to an absorbed EM
quantum, and that with the minus sign to an emitted one.

Under the effect of the time-dependent perturbation given by
Eq.~(\ref{V}), in a time $t$ the system evolves from state $m$ to
state $n$ according to the first order transition
amplitude~\cite{Landau3}
\begin{align}
  \label{anm.mode}
  a^\pm_{n,m}(\bm{k},t) = -\frac{\I}{\hbar} \int_{0}^{t} \D{s}\
  V^\pm_{n,m}(\bm{k},s) e^{\frac{\I}{\hbar} (E_n-E_m)s},
\end{align}
where
\begin{align}
  \label{Vnm.mode}
  V^\pm_{n,m}(\bm{k},s) &= \bra{E_n}\hat{V}^\pm(\bm{k},s) \ket{E_m}
  \nonumber \\
  &= -\frac{\mu}{2} \sum_{i=1}^{N} \sum_{h=1}^{3} \bra{E_n}
  \sigma_i^{h} \ket{E_m} B_{h}(\bm{k})
  e^{\pm\I(\bm{k}\cdot\bm{r}_i-kcs)} .
\end{align}
The squared modulus of Eq.~(\ref{anm.mode}) gives the probability of
the system's evolving in a time $t$ from state $m$ to state $n$ due to
the interaction with the mode $(\bm{k},\pm)$.  However, for $m$ and
$n$ fixed, there are several modes $(\bm{k},\pm)$ contributing to the
transition $m\to n$, namely, all those compatible with the energy
conservation law $E_n=E_m\pm \hbar k c$.  We thus evaluate the
effective probability for the transition $m\to n$ in a time $t$ by
taking the expectation of $\abs{a^\pm_{n,m}(\bm{k},t)}^2$ over all the
modes in the cavity:
\begin{align}
  \label{anm2.ave}
  &\E{\abs{a^\pm_{n,m}(\bm{k},t)}^2} \nonumber \\
  &\quad = \frac{\mu^2}{4\hbar^2} \int_{0}^{t}\D{s}\
  e^{\frac{\I}{\hbar} (E_n-E_m)s} \int_{0}^{t}\D{u}\
  e^{-\frac{\I}{\hbar} (E_n-E_m)u} \nonumber \\ &\quad\quad\times
  \sum_{i=1}^{N} \sum_{j=1}^{N} \sum_{h=1}^{3} \sum_{l=1}^{3}
  \bra{E_n} \sigma_i^{h} \ket{E_m} \overline{\bra{E_n} \sigma_j^{l}
    \ket{E_m}} \nonumber \\ &\quad\quad\times \E{B_{h}(\bm{k})
    e^{\pm\I(\bm{k}\cdot\bm{r}_i-kcs)} B_{l}(\bm{k})
    e^{\mp\I(\bm{k}\cdot\bm{r}_j-kcu)}}.
\end{align}
In the above formula, all the statistical properties of the blackbody
radiation are enclosed in the field-field correlation function $
\E{B_{h}(\bm{k}) e^{\pm\I(\bm{k}\cdot\bm{r}_i-kcs)} B_{l}(\bm{k})
  e^{\mp\I(\bm{k}\cdot\bm{r}_j-kcu)}} $.  This correlation function
was first evaluated by Bourret~\cite{Bourret1960} in the case of real
fields and then extended to the case of complex fields by Kano and
Wolf~\cite{KanoWolf1962} and by Metha and Wolf~\cite{MethaWolf1964I}.
The result which applies directly to our case is~\cite{MethaWolf1964I}
\begin{align}
  \label{cf}
  &\E{B_{h}(\bm{k}) e^{\pm\I(\bm{k}\cdot\bm{r}_i-kcs)} B_{l}(\bm{k})
    e^{\mp\I(\bm{k}\cdot\bm{r}_j-kcu)}} \nonumber \\ &\quad= \int
  \D{\bm{k}}\ e^{\pm\I(\bm{k}\cdot(\bm{r}_i-\bm{r}_j)-kc(s-u))}
  \nonumber \\ &\quad\quad\times \frac{1}{\pi^2} \frac{\hbar
    kc}{e^{\hbar kc/k_B T}-1} \left(\delta_{h,l}-\frac{k_h
      k_l}{k^2}\right).
\end{align}
Note that an identical expression holds for the electric-field
correlation function.  On plugging Eq.~(\ref{cf}) into
Eq.~(\ref{anm2.ave}), the integrals over the times $s$ and $u$ can be
separately performed as follows
\begin{align}
  &\int_{0}^{t}\D{s}\ e^{\frac{\I}{\hbar} (E_n-E_m\mp\hbar kc)s}
  \int_{0}^{t}\D{u}\ e^{-\frac{\I}{\hbar} (E_n-E_m\mp\hbar kc)u}
  \nonumber \\ &\quad= \abs{\int_{0}^{t}\D{s}\ e^{\frac{\I}{\hbar}
      (E_n-E_m\mp\hbar kc)s}}^2 \nonumber \\ &\quad= 2\pi\hbar t\
  \frac{ \sin^2\left[(E_n-E_m\mp\hbar kc) \frac{t}{2\hbar}\right]}
  {\pi (E_n-E_m\mp\hbar kc)^2 \frac{t}{2\hbar}} \nonumber \\ &\quad
  \simeq 2\pi\hbar t\ \delta(E_n-E_m\mp\hbar kc).
\end{align}
As usual, this approximation is proved to be accurate for $t$ large by
using the representation of the Dirac distribution
\begin{align}
  \delta(x) = \lim_{y\to\infty} \frac{\sin^2 (xy)}{\pi x^2 y}.
\end{align}
We conclude that the effective transition rate from state $m$ to state
$n$ is
\begin{align}
  \label{Pnm}
  P^\pm_{n,m} &= \frac{1}{t} \E{\abs{a^\pm_{n,m}(\bm{k},t)}^2}
  \nonumber \\ &= \frac{\mu^2}{2\pi\hbar} \sum_{i=1}^{N}
  \sum_{j=1}^{N} \sum_{h=1}^{3} \sum_{l=1}^{3} \bra{E_n} \sigma_i^{h}
  \ket{E_m} \overline{\bra{E_n} \sigma_j^{l} \ket{E_m}} \nonumber \\
  &\quad\times \int \D{\bm{k}}\
  e^{\pm\I\bm{k}\cdot(\bm{r}_i-\bm{r}_j)} \frac{\hbar kc}{e^{\hbar
      kc/k_B T}-1} \nonumber \\ &\quad\times \delta(E_n-E_m\mp\hbar
  kc) \left(\delta_{h,l}-\frac{k_h k_l}{k^2}\right).
\end{align}
The plus-minus sign in the factor
$e^{\pm\I\bm{k}\cdot(\bm{r}_i-\bm{r}_j)}$ is irrelevant and is omitted
hereafter.

In Eq.~(\ref{Pnm}) we can evaluate the integral over the modulus $k$
of the wave vector by means of the Dirac distribution.  We write
$\bm{k}=k \bm{u}$, with $k=\abs{\bm{k}}$ and
$\bm{u}=(\sin\theta\cos\phi,\sin\theta\sin\phi,\cos\theta)$ unit
vector given in terms of the longitudinal and azimuthal angles
$\theta$ and $\phi$ ranging, respectively, in $[0,\pi]$ and
$[0,2\pi]$.  Using $\D{\bm{k}} = k^2 \D{k} \sin\theta\D{\theta}
\D{\phi}$, we get
\begin{align}
  \label{Pnm.general}
  P^\pm_{n,m} &= \frac{\mu^2}{2\pi\hbar c^3}\
  \frac{\omega_{n,m}^3}{e^{\hbar\omega_{n,m}/k_B T}-1} \nonumber \\
  &\quad\times \sum_{i=1}^{N} \sum_{j=1}^{N} \sum_{h=1}^{3}
  \sum_{l=1}^{3} Q_{n,m}^{i,j;h,l} \nonumber \\ &\quad\times \bra{E_n}
  \sigma_i^{h} \ket{E_m} \overline{\bra{E_n} \sigma_j^{l} \ket{E_m}} ,
\end{align}
where
\begin{align}
  \label{omeganm}
  \omega_{n,m}=\abs{E_n-E_m}/\hbar,
\end{align}
and
\begin{align}
  \label{Qnm}
  Q_{n,m}^{i,j;h,l} = \int_{0}^{\pi} \!\!  \sin\theta \D{\theta}
  \int_{0}^{2\pi} \!\!\! \D{\phi}\
  e^{\I\bm{u}\cdot(\bm{r}_i-\bm{r}_j)\omega_{n,m}/c}
  \left(\delta_{h,l}-u_hu_l\right).
\end{align}
The notation in Eq.~(\ref{Pnm.general}) has been simplified by using
$\omega_{n,m}=\abs{E_n-E_m}/\hbar$ instead of two separate angular
frequencies $\omega^\pm_{n,m}$ for the energy-gaining and the
energy-losing transitions. Actually, it results that $\omega^\pm_{n,m}
= \pm (E_n-E_m)/\hbar=\abs{E_n-E_m}/\hbar$.  Note that for $E_n=E_m$
we have $P^\pm_{n,m}=0$, which expresses the fact that there is no
zero-mode (constant) EM field, in agreement with the homogeneity and
isotropy of the radiation in the cavity.  Contributions in which
$E_m=E_n$, including also the case $m=n$, may appear only at higher
orders of the time-dependent perturbation theory.

Equation~(\ref{Pnm.general}) is our general expression of the
transition rate $m \to n$ for a system of $N$ magnetic dipoles
interacting with blackbody radiation. In the case of $N$ electric
dipoles, we have an identical formula with $\mu \bm{\sigma}_i$
replaced by $\bm{p}_i$, the moment of the $i$th electric dipole.  In
this case, as well as in the case of spatial magnetic dipoles,
Eq.~(\ref{Pnm.general}) still holds if $d_{n,m}~ \omega_{n,m}/c \ll
1$, where $d_{n,m}= \max_i \left| \bra{E_n} \delta
  \bm{r}_i\ket{E_m}\right|$ and $\delta \bm{r}_i$ is the vector
between the positive and negative charges of the $i$th dipole. This
condition allows for a long-wavelength approximation in
Eq.~(\ref{Vnm.mode}), so that the phase factors
$e^{\pm\I(\bm{k}\cdot\bm{r}_i-kcs)}$ can still be considered constant
factors in respect of the $N$-body matrix element.

Depending on the spatial distribution of the $N$ dipoles and the value
of $\omega_{n,m}$, two limiting regimes of Eq.~(\ref{Pnm.general}) can
be attained.

\textit{Fully coherent limit.}  If the $N$ dipoles are localized in a
region of extension $\ell \ll \lambda_{n,m}$, where
$\lambda_{n,m}=2\pi c/\omega_{n,m}$, we have
$\abs{\bm{r}_i-\bm{r}_j}\omega_{n,m}/c \ll 2\pi$ for any $i,j$.  This
implies that in Eq.~(\ref{Qnm}) we can approximate
$e^{\I\bm{u}\cdot(\bm{r}_i-\bm{r}_j)\omega_{n,m}/c} \simeq 1$ and
straightforwardly perform the integrals over $\theta$ and $\phi$. The
result is
\begin{align}
  \label{Qnm.coherent}
  Q_{n,m}^{i,j;h,l} &= \int_{0}^{\pi} \!\!  \sin\theta \D{\theta}
  \int_{0}^{2\pi} \!\!\! \D{\phi}\ \left(\delta_{h,l}-u_hu_l\right)
  \nonumber \\ &= \frac{8\pi}{3}~\delta_{h,l}
\end{align}
In this limit, Eq.~(\ref{Pnm.general}) reduces to
\begin{align}
  \label{Pnm.coherent}
  P^\pm_{n,m} &= \frac{4\mu^2}{3\hbar c^3}\
  \frac{\omega_{n,m}^3}{e^{\hbar\omega_{n,m}/k_B T}-1} \nonumber \\
  &\quad\times \sum_{h=1}^{3} \abs{\bra{E_n} \sum_{i=1}^{N}
    \sigma_i^{h} \ket{E_m}}^2 .
\end{align}
The total dipole of the system, $\sum_{i=1}^{N} \mu\bm{\sigma}_i$,
couples coherently with an isotropic radiation mode of angular
frequency $\omega_{n,m}$ weighed according to the Plank distribution.
In particular, in the limit of non interacting particles we get
$\sum_m P^\pm_{n,m}=O(N)$.

For $N=1$, Eq.~(\ref{Pnm.coherent}) is the standard textbook formula
based on the long-wavelength approximation. Reference~\cite{Davydov}
suggests that this formula is appropriate for describing many electron
atoms in a blackbody radiation upon just replacing the dipole
(electric or magnetic) of the single electron with the total dipole of
the electrons in the atom.

\textit{Fully incoherent limit.}  A much different result is obtained
if the $N$ dipoles are separated from each other by a distance much
longer than $\lambda_{n,m}$. Suppose, for simplicity, that the dipoles
occupy the sites of a regular linear lattice of spacing $a$.  Choosing
the reference frame in such a way that the lattice points are
determined by the vectors $\bm{r}_i = (0,0,ai)$, we have
$\bm{u}\cdot(\bm{r}_i-\bm{r}_j)=(i-j)a\cos\theta$, with
$i,j=1,\dots,N$.  In the general expression for the coefficients
$Q_{n,m}^{i,j;h,l}$ given by Eq.~(\ref{Qnm}), we can separately
evaluate the integral over $\phi$ and obtain
\begin{align}
  \int_{0}^{2\pi} \!\!\! \D{\phi}\ \left(\delta_{h,l}-u_hu_l\right) =
  f_h(\theta)\ \delta_{h,l},
\end{align}
where
\begin{align}
  f_h(\theta) = \left\{
    \begin{array}{ll}
      2\pi-\pi \sin^2\theta &\quad h=1,2 
      \\
      2\pi-2 \pi \cos^2\theta &\quad h=3
    \end{array}
  \right. .
\end{align}
Performing the remaining integral over $\theta$, we get
\begin{align}
  \label{Qnm.incoherent.1}
  Q_{n,m}^{i,j;h,l} &= \int_{0}^{\pi} \!\!  \sin\theta \D{\theta}\
  e^{\I a \cos \theta (i-j)\omega_{n,m}/c}\ f_h(\theta)\ \delta_{h,l}
  \nonumber \\ &= q_{n,m}^{i,j;h}\ \delta_{h,l},
\end{align}
where
\begin{align*}
  q_{n,m}^{i,j;h} = \left\{
    \begin{array}{ll}
      \frac{4\pi\left(b_{n,m}^{i,j}\cos b_{n,m}^{i,j} 
          + ((b_{n,m}^{i,j})^2-1)\sin b_{n,m}^{i,j} \right)}{(b_{n,m}^{i,j})^3}
      &\mbox{$i \neq j$, $h=1,2$} 
      \\
      \frac{8\pi \left(\sin b_{n,m}^{i,j} - 
          b_{n,m}^{i,j}\cos b_{n,m}^{i,j} \right)}{(b_{n,m}^{i,j})^3}
      &\mbox{$i \neq j$, $h=3$} 
      \\
      \frac{8\pi}{3} &\mbox{$i=j$}
    \end{array}
  \right.
\end{align*}
and
\begin{align}
  b_{n,m}^{i,j} = (i-j)a\omega_{n,m}/c.
\end{align}
For $a\omega_{n,m}/c \gg 2\pi$, i.e., $a \gg \lambda_{n,m}$,
neglecting terms $O(\lambda_{n,m}/a)$, we can approximate
\begin{align}
  \label{Qnm.incoherent.2}
  Q_{n,m}^{i,j;h,l} = \frac{8\pi}{3}\ \delta_{i,j} \delta_{h,l}.
\end{align}
In this limit, Eq.~(\ref{Pnm.general}) reduces to
\begin{align}
  \label{Pnm.incoherent}
  P^\pm_{n,m} &= \frac{4\mu^2}{3\hbar c^3}\
  \frac{\omega_{n,m}^3}{e^{\hbar\omega_{n,m}/k_B T}-1} \nonumber \\
  &\quad\times \sum_{i=1}^{N} \sum_{h=1}^{3} \abs{\bra{E_n}
    \sigma_i^{h} \ket{E_m}}^2 .
\end{align}
The transition rate $m\to n$ is now the incoherent sum of $N$
contributions from the single dipoles. Note, however, that the matrix
elements $\bra{E_n} \sigma_i^{h} \ket{E_m}$ between two eigenstates of
$\hat{H}$ still retain their full $N$-body character.
Equation~(\ref{Qnm.incoherent.2}) and, therefore, the fully incoherent
formula (\ref{Pnm.incoherent}), apply also when the $N$ dipoles are
placed at arbitrary positions, provided the minimal distance between
two of them is still $a \gg \lambda_{n,m}$.  As in the coherent case,
for non interacting particles we have, again, $\sum_m P^\pm_{n,m}
=O(N)$.

The conditions for the validity of the fully coherent and fully
incoherent limits are better expressed in terms of the energies of the
levels $n$ and $m$. We have, respectively,
\begin{align}
  \label{coherent}
  \abs{E_n-E_m} \ll hc/\ell, \qquad \ell=\max_{i\neq j}
  \abs{\bm{r}_i-\bm{r}_j},
\end{align}
\begin{align}
  \label{incoherent}
  \abs{E_n-E_m} \gg hc/a, \qquad a=\min_{i\neq j}
  \abs{\bm{r}_i-\bm{r}_j}.
\end{align}
Observing that $hc = 1.23\ \mbox{eV $\mu$m}$, it is evident that for
atomic or molecular systems in which $\abs{E_n-E_m}$ is, at most, a
few electron volts and $\ell$ is not larger than a few tens of
angstroms, Eq.~(\ref{coherent}) is well satisfied and the fully
coherent formula (\ref{Pnm.coherent}) applies.  Viceversa, for
microscopic systems in which $a$ is 1 $\mu\mbox{m}$ and the
energy-level separations $\abs{E_n-E_m}$ are much larger than the
atomic electron volt scale, condition~(\ref{incoherent}) is met and we
can apply the fully incoherent formula, (\ref{Pnm.incoherent}).
However, this may not be true for systems having, in the thermodynamic
limit $N\to\infty$, a phase transition which implies the existence of
a vanishing gap. For systems of intermediate extension or in
particular regions of the energy spectrum in the presence of a phase
transitions, the general formula, (\ref{Pnm.general}), must be
applied.

Equation~(\ref{Pnm.general}) has been obtained on the basis of the
semiclassical theory of radiation. The field-field correlation
function, (\ref{cf}), which is its foundation, can be evaluated in the
framework of the quantized theory of radiation and provides an
identical result~\cite{MethaWolf1964II}. In this case, however, the
interaction of the $N$-body system with the zero-point energy of the
quantized EM modes gives rise to spontaneous emission processes which
add to the transition rate for stimulated emission $P^{-}_{n,m}$. The
total emission rate, stimulated and spontaneous, is still given by our
$P^{-}_{n,m}$, with the average number of photons at energy
$\hbar\omega_{n,m}$ increased by one unity \cite{Davydov}, namely,
\begin{align*}
  \frac{1}{e^{\hbar\omega_{n,m}/k_B T}-1} \to
  \frac{1}{e^{\hbar\omega_{n,m}/k_B T}-1}+1 .
\end{align*}

% \bibliography{goldenrule} % run pdflatex bibtex pdfltex pdflatex
% \input{goldenrule.bbl} % to be used before submission

% merlin.mbs apsrev4-1.bst 2010-07-25 4.21a (PWD, AO, DPC) hacked
% Control: key (0) Control: author (0) dotless jnrlst Control: editor
% formatted (1) identically to author Control: production of article
% title (0) allowed Control: page (1) range Control: year (0) verbatim
% Control: production of eprint (0) enabled
%

\end{document}